\documentclass[universe,article,accept,pdftex,moreauthors]{Definitions/mdpi} 

\usepackage{comment}

\newcommand{\nat}{Nature}

\firstpage{1} 
\pubvolume{1}
\issuenum{1}
\articlenumber{0}
\pubyear{2024}
\copyrightyear{2024}
\datereceived{13 December 2023} 
\daterevised{2 January 2024} 
\dateaccepted{8 January 2024} 
\datepublished{ } 
\hreflink{https://doi.org/} 

\Title{Covering Factor of the Dust-Driven Broad-Line Region Clouds}

\TitleCitation{Covering Factor of the Dust-Driven Broad-Line Region Clouds}

\Author{
Mohammad-Hassan Naddaf $^{1,2,}$*$^{,\dagger}$\orcidA{}
and Bo\.zena Czerny
$^{1,\dagger}$\orcidB{}}

\AuthorNames{Mohammad Hassan Naddaf, and Bo\.zena Czerny}

\AuthorCitation{Naddaf, M.H.; Czerny, B.}

\address{%
$^{1}$ \quad Center for Theoretical Physics, Polish Academy of Sciences, Lotnikow 32/46, 02-668 Warsaw, Poland; bcz@cft.edu.pl\\
$^{2}$ \quad Institut d’Astrophysique et de Géophysique, Université de Liège, Allée du six août 19c, B-4000~Liège~(Sart-Tilman), Belgium}

\corres{Correspondence: naddaf@cft.edu.pl}

\firstnote{These authors contributed equally to this work.} 

\abstract{The origin of the broad-line region (BLR) clouds in active galactic nuclei is still under discussion. We develop a scenario in which the clouds in the outer, less ionized part of the BLR are launched by the radiation pressure acting on dust. Most of the outflow forms a failed wind, so we refer to it as failed radiatively accelerated dusty outflow (FRADO), but, for a certain parameter range, actual outflow also takes place. We aim to test the model predictions. In this paper, we present the calculation of the angular distribution of clouds and the net covering factor as this affects the fraction of radiation that can be intercepted and reprocessed in the form of the H$\beta$ or Mg II emission line.  
The results reveal that the covering factor is intricately linked to the mass, accretion rate, and metallicity of the clouds. Notably, as these parameters increase, so does the covering factor, shedding light on the dynamic interplay between the central engine and the surrounding material in AGNs.}


\begin{document}


\section{Introduction}

An active galactic nucleus (AGN) is the intensely energetic and compact central region of a galaxy, characterized by a supermassive black hole actively accreting mass. This accretion process releases an extraordinary amount of energy, leading to the emission of various forms of radiation across the electromagnetic spectrum. AGNs exhibit diverse phenomena, including strong and variable emission in optical, X-ray, and radio wavelengths. The unified model of AGNs seeks to explain the observed diversity in AGN properties by considering the orientation of the observer relative to the accretion disk and surrounding structures, such as the torus of gas and dust \citep{antonucci1993, netzer2015, Almeida2017}.

The covering factor, which denotes the proportion of the sky obscured by the obscuring material when viewed from the accreting supermassive black hole (SMBH), plays a pivotal role in governing the intensity of reprocessed X-ray and infrared (IR) radiation \citep{lawrence2010}. Over the past decade, investigations into the covering factors of diverse samples of active galactic nuclei (AGNs) at various wavelengths have unveiled trends correlated with luminosity and redshift. Determining the covering factor involves spectral modeling in both the IR and X-ray spectra, as expounded in the following section. Two additional commonly employed methods are as follows. First, in the IR, the ratio between mid-infrared (MIR) and AGN bolometric luminosity serves as a proxy for the torus reprocessing efficiency. The fraction of optical/UV and X-ray radiation reprocessed by the torus and observed in the MIR is directly proportional to its covering factor \citep{Granato1994, Elitzur2008, Lanz2019}. Second, in the X-ray, the covering factor of the gas and dust enveloping the SMBH can be estimated through statistical analyses by examining the absorption characteristics of extensive AGN samples \citep{Ueda2003, ricci2015}. Due to the compact nature of the X-ray corona, the column density obtained from the X-ray spectroscopy of individual objects offers information solely along a specific line of sight. Analyzing large sets of objects enables the exploration of random inclination angles, thus enhancing our comprehension of the average attributes of the obscuring material. Notably, the likelihood of observing an obscured AGN is directly proportional to the covering factor of the surrounding gas and dust.

In this paper, we calculate the covering factor of broad-line region clouds that are produced due to the radiatively dust-driven mechanism---more specifically, the FRADO model. In contrast to most models that adopt the BLR with general but arbitrary geometries \citep{netzer1993, ward2014, pancoast2014, tek2016, tek2018}, this model possesses predictive power. The covering factor is not assumed but rather results from the global source parameters such as the black hole mass and accretion rate. Therefore, it is crucial to test the validity of these predictions. The outline of the paper is as follows. In Section \ref{sec:FRADOmodel}, we briefly introduce the FRADO model and its implications for the covering factor. We then discuss the results of the distribution of clouds based on the FRADO model, and, subsequently, the calculation of the covering factor of clouds is presented in Section \ref{sec:results}. In the last two sections, we present a discussion of the results and the conclusions.

\section{FRADO Model}\label{sec:FRADOmodel}

The analytical FRADO model, crafted a decade ago \citep{czerny2011, czerny2015, czerny2016, czerny2017}, explores the dynamics of dusty material raised from the accretion disk surface under the influence of radiation pressure acting on dust present in the surface layers of the outer part of the accretion disk. The underlying accretion disk is based on the standard model of Shakura--Sunyaev~\citep{SS73}, which works appropriately for the range of accretion rates approximately from $0.01$ up to almost $1$ Eddington \citep{Mineshige1995, Esin1997, Esin1998}. The scenario aims to address the issue of the physical origin of the BLR clouds in the outer region, where low ionization lines like H$\beta$ and Mg II originate. It takes the form of a  failed wind, depicted in a one-dimensional manner---specifically, in the vertical direction—without accounting for orbital motion. In this model, the clumps of material {(dusty clouds)}, consisting of dust and gas at the surface layers of the cold accretion disk \citep{rees1969, dong2008}, are propelled by the local radiation flux emanating from the accretion disk itself. Once these clumps attain high altitudes, intense irradiation from the central disk leads to the loss of their dust content. The residual gaseous clumps {(dustless clouds)} then undergo ballistic motion within the gravitational field of the central black hole before ultimately returning to the disk surface.  The scenario has considerable predictive power, offering a means to test its predictions by simply following the model consequences, without including new parameters. The FRADO model was subsequently developed by following the motion in 3D and applying the careful computation of the radiation pressure force \citep{naddaf2021}, and its predictions were considered for the cloud impact on the disk \citep{muller2022} and probability to observe the broad absorption line (BAL; \citep{BAL_2023}). Here, we address the problem of the covering factor resulting from the scenario.

 We use here the basic model as developed by \citet{naddaf2021}. The underlying accretion disk is modeled using the basic idea of \citet{SS73}, which predicts the effective temperature of the stationary accretion disk as a function of the radius in the Newtonian approximation of the gravitational field. This is not accurate in the innermost part of the disk, where relativistic effects and the black hole spin are important, but it well describes the outer parts of the disk. We assume the local black body emission without any color correction to the local spectrum. The innermost part of the disk can be additionally modified by the presence of the hot and warm corona (e.g., \citep{kubota2018}). However, in dust driving, the effect of the shielding is important, and, in our model, this shielding is height-dependent, with the fraction of the disk visible to the cloud rising with the cloud height. Therefore, the radiation from the innermost part does not affect the cloud launching, as this is achieved by the predominantly local radiation, although it might affect the final velocity of the escaping clouds. However, the launching depends essentially on the disk height, and this in turn requires the proper description of the disk vertical cooling through opacities. We use, for this purpose, the code of \citet{rozanska1999}, which contains the effect of electron scattering and Kramer opacities for a partially ionized medium, including the molecules and dust grains.

Clouds are launched from the disk surface under the effect of the radiation acting on dust. Line driving (as done in the QWIND model of \citep{risaliti2010}) is not included since the combination of line and dust driving is very complex. Dust driving requires non-local effects from integration over the spectra from the part of the disk visible to the cloud at each radius and height, while line driving couples the radiative force with the velocity due to the use of the force multiplier. Dust opacities are modeled, including the wavelength-dependent cross-sections for each grain size, and the dust properties are based on the standard MNR model of \citet{mathis1977}. Launched clouds preserve the angular momentum of the Keplerian orbit at the launch point along the entire trajectory. The full model is parameterized by the value of the black hole mass, $M_{BH}$; the dimensionless accretion rate $\dot m$ in units of the Eddington critical accretion rate; and the metallicity $Z$ of the material in units of the solar metallicity. 
For more details on the description of the model and results, see~\citep{naddaf2021, Naddaf2022}. In this paper, we consider a grid of models corresponding to $M_{BH}$ of $10^{8} M_{\odot}$, $\dot m$ of 0.1 and 1 in Eddington units, and $Z$ of 1 and 5 in solar units.

\begin{figure}[H]
\begin{adjustwidth}{-\extralength}{0cm}
\centering
\includegraphics[width=17 cm]{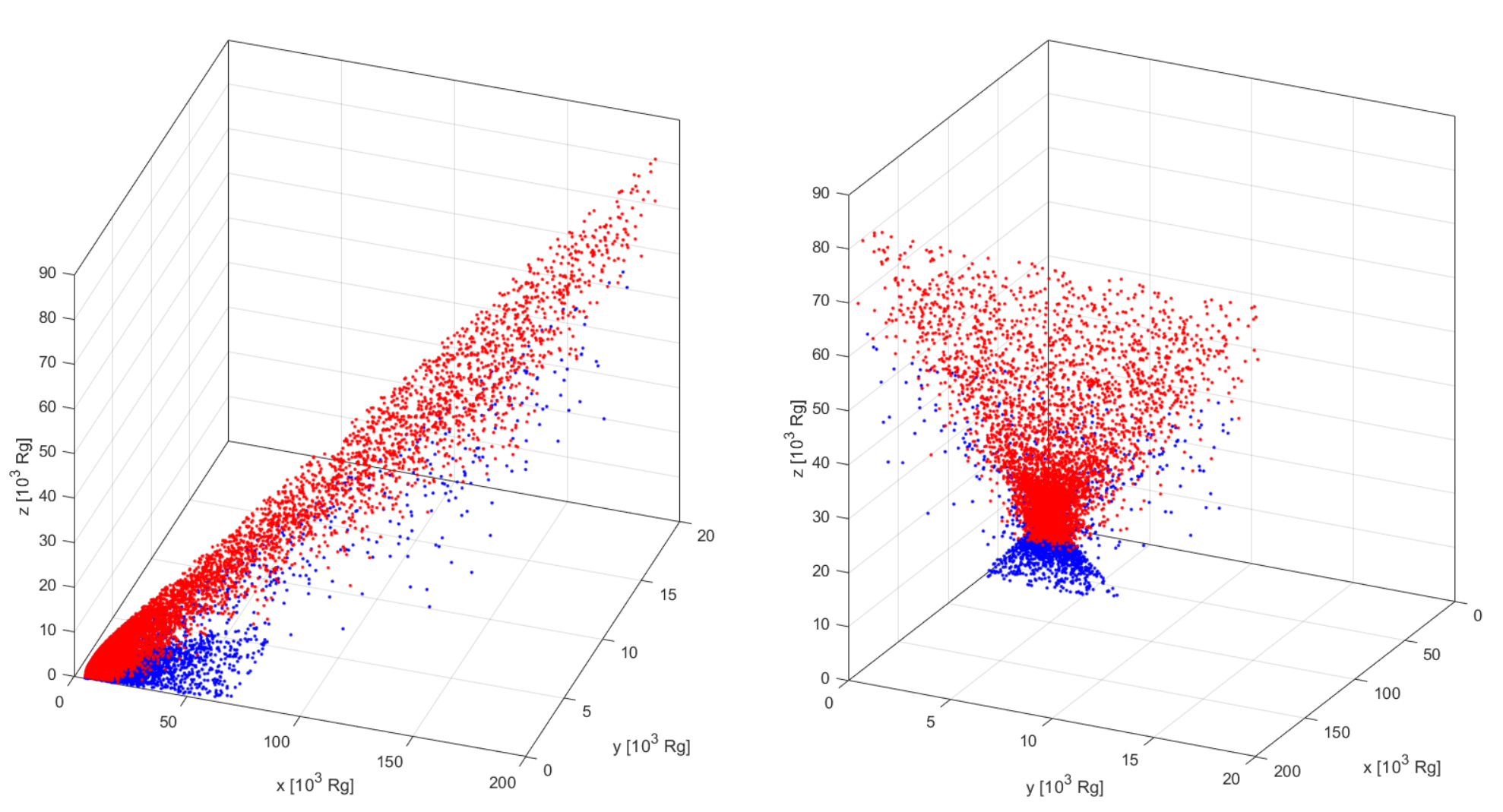}
\end{adjustwidth}
\caption{Two perspectives of the same slice of distribution of particles (clouds) based on FRADO, for a case at Eddington rate with the black hole mass of $10^{8} M_{\odot}$ and metallicity of 5 times solar. {$R_{g}$ is the gravitational radius of the black hole of $10^{8} M_{\odot}$}. Dusty and dustless clouds are marked in blue and red, respectively.\label{fig:Z5M8m100Distribution}}
\end{figure}

Figure \ref{fig:Z5M8m100Distribution} shows a slice of the azimuthally symmetric distribution of radiatively dust-driven BLR clouds in the FRADO model for an arbitrary case with a black hole mass of $10^{8} M_{\odot}$ and metallicity of 5 times solar, which is at the Eddington rate.

\subsection{Cloud Parameters and the Outflow Net Flux}

The computation of the differential and total covering factor requires additional specifications that do not come directly from the model of the dynamics. We address this issue in \citet{muller2022}, where we examine the consequences of the impact of returning clouds on the disk,  which results in gamma-ray emission. We follow a similar path here; thus, we consider two values of the column density of a cloud, $N_{H}=$  $10^{23}$ and $10^{24}$ cm$^{-2}$, and two values of the cloud local density,  $n_{c} = 10^{11}$ and $10^{12}$ cm$^{-3}$. These values well represent the expected range for the BLR material. For the column density, the lower and upper values correspond to BLR clouds being optically thin or thick, respectively~\citep{muller2022}. In addition, they also well represent the range of the column densities of individual clouds as they form as the result of the thermal instability characteristic for the irradiated medium \citep{krolik1981,begelman1990}. The range of densities is motivated by the most recent observational estimates (e.g., \citep{panda2018}), and by theoretical models of clumpiness based on radiation pressure confinement \citep{baskin2018}. The normalization of the total number of clouds is now performed differently, rather following \citet{naddaf2022pas} and \citet{Naddaf_2022Dynam}, and we concentrate on solutions that show outflowing clouds supplementing the failed part of the wind. The number of clouds at each trajectory in the model itself represents the time spent in completing the path, but the cloud sampling density is arbitrary and each cloud in the computation has the meaning of the effective cloud. Thus, the total number of clouds is then renormalized to the outflow rate, which represents the total mass of clouds (after scaling) and the timescale of the outflow:
\begin{equation}
\dot M_{\rm out} = \dfrac{M_{\rm clouds_{total}}} {\tau_{\rm escape}}.
\label{eq:scaling}
\end{equation}

This method is a simplification since it includes a mean timescale of the BLR replacement. Direct computations of the actual number of clouds on each trajectory would be numerically too time-consuming as this factor in various model parameters ranges from 100 to $\sim$$10^5$. 


\subsection{Differential and Total Covering Factor}
\label{sect:DCF}

We are interested in the total covering factor since it normalizes the contribution of the BLR to the total spectrum, and this contribution must correspond to the observed spectral features in the optical/UV band---in particular, the equivalent width (EW) of the lines. However,  we are also interested in the differential covering factor since it gives statistical information about the probability of having a single BLR cloud temporarily located along the line of sight towards the observer. Such events were observed in the past as transits lasting a few days in the X-ray band (e.g., \citep{markowitz2014}), so this also offers a test of the model. In our previous paper addressing the broad absorption line (BAL) phenomenon, we simply treated the whole outflow cone as a solid body \citep{BAL_2023}. Now, we consider the 3D cloud distribution, so we find it useful to introduce the concept of the differential covering factor,  DCF(i), which gives the fraction of the sky covered by clouds at a fixed viewing angle towards the observer. We use a linear scale in the viewing angle, with 90 steps between 0 and 90 deg (measured from the symmetry axis of $z$), and we count the clouds at each sky ring separately. We simply count the solid angles of each cloud, without the effect of overlapping, and when the clouds fill the ring completely (in computations, it appears as effective coverage larger than 1), we assume that it saturates at 1.

The total (global) covering factor $C_{\rm global}$ is the fraction of the sky covered by clouds when integrating over all viewing angles. It is normalized to the solid angle of the half-sphere. 

The method using Equation~\eqref{eq:scaling} applies to solutions that contain an escaping stream of clouds. However, as we showed in \citet{naddaf2021}, at low masses and accretion rates, only a failed wind forms. In these cases, we cannot calculate the scaling factor, but we know that such clouds cover the available space densely, so, in such a case, we calculate the covering factor geometrically, with DCF rapidly changing from 0 (no clouds) to 1 (full coverage at a larger angle).

\section{Results}\label{sec:results}

We first analyze in detail the model  with a black hole mass of $10^{8} M_{\odot}$, solar metallicity, and the flow luminosity corresponding to the Eddington accretion rate. The  outflow rate (as in Equation~\eqref{eq:scaling}) calculated in \citet{Naddaf_2022Dynam} for these parameters gives $10^{-4} \dot M_{accr}$ or $3 \times 10^{-2} \dot M_{accr}$ (where $\dot M_{accr}$ is the accretion rate of the source) for optically thin or optically thick wind launching, which corresponds to $2 \times 10^{-4} M_{\odot}$ yr$^{-1}$ or $6 \times 10^{-2} M_{\odot}$~yr$^{-1}$, respectively. The resolution of the computations gives us directly 866,664  clouds (total mass  of $9.7 \times 10^{-4} M_{\odot}$ for the column density and density of $10^{24}$ cm$^{-2}$ and $10^{12}$ cm$^{-3}$, respectively). The escape timescale in this model is 1000 years. Therefore, each cloud in the computations represents 100 physical clouds, and $2 \times 10^4$ clouds, respectively. We use the corresponding factor when plotting the differential covering factor and giving the total covering factor, as defined in Section~\ref{sect:DCF}.

\begin{figure}[H]
\includegraphics[width=7 cm]{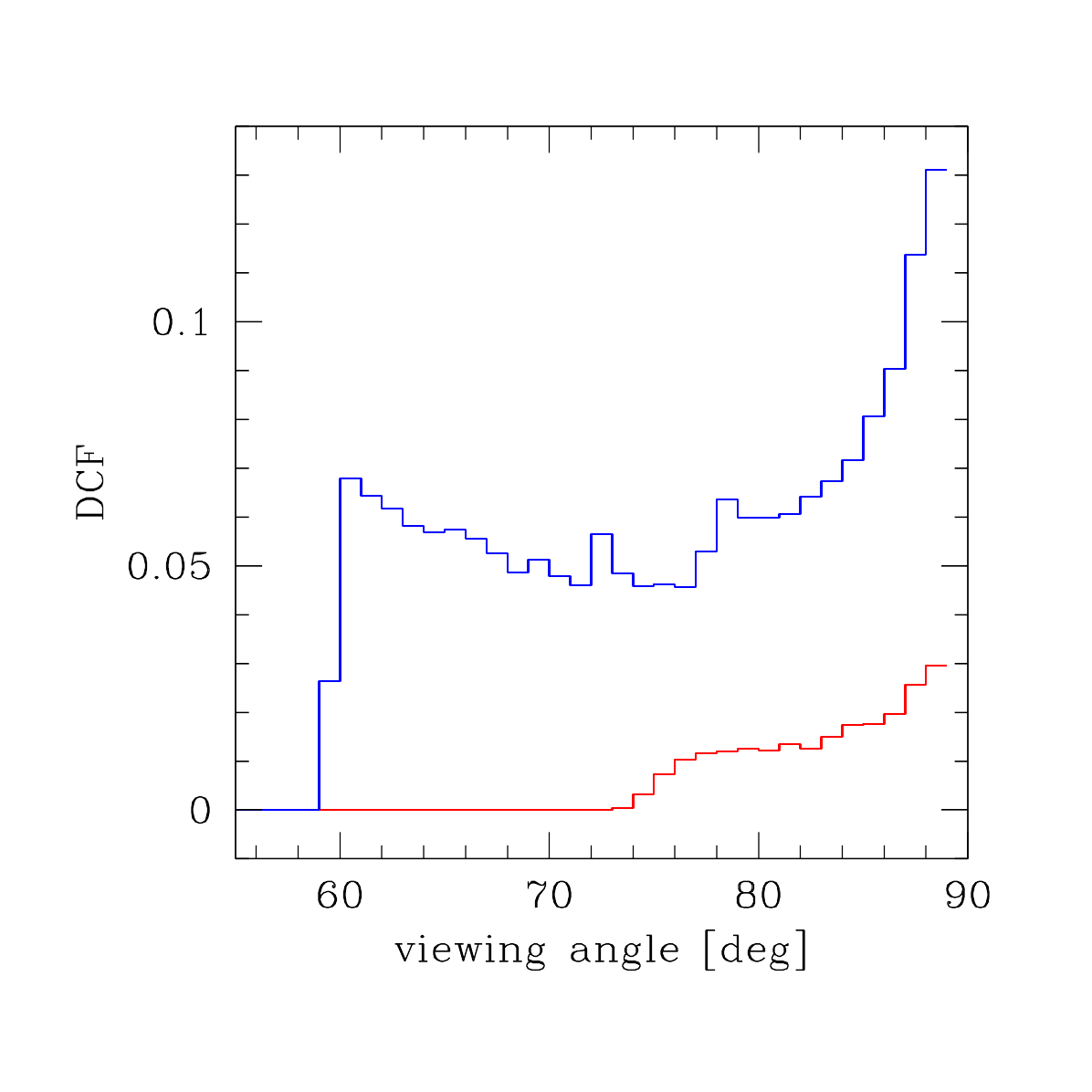}
\includegraphics[width=7 cm]{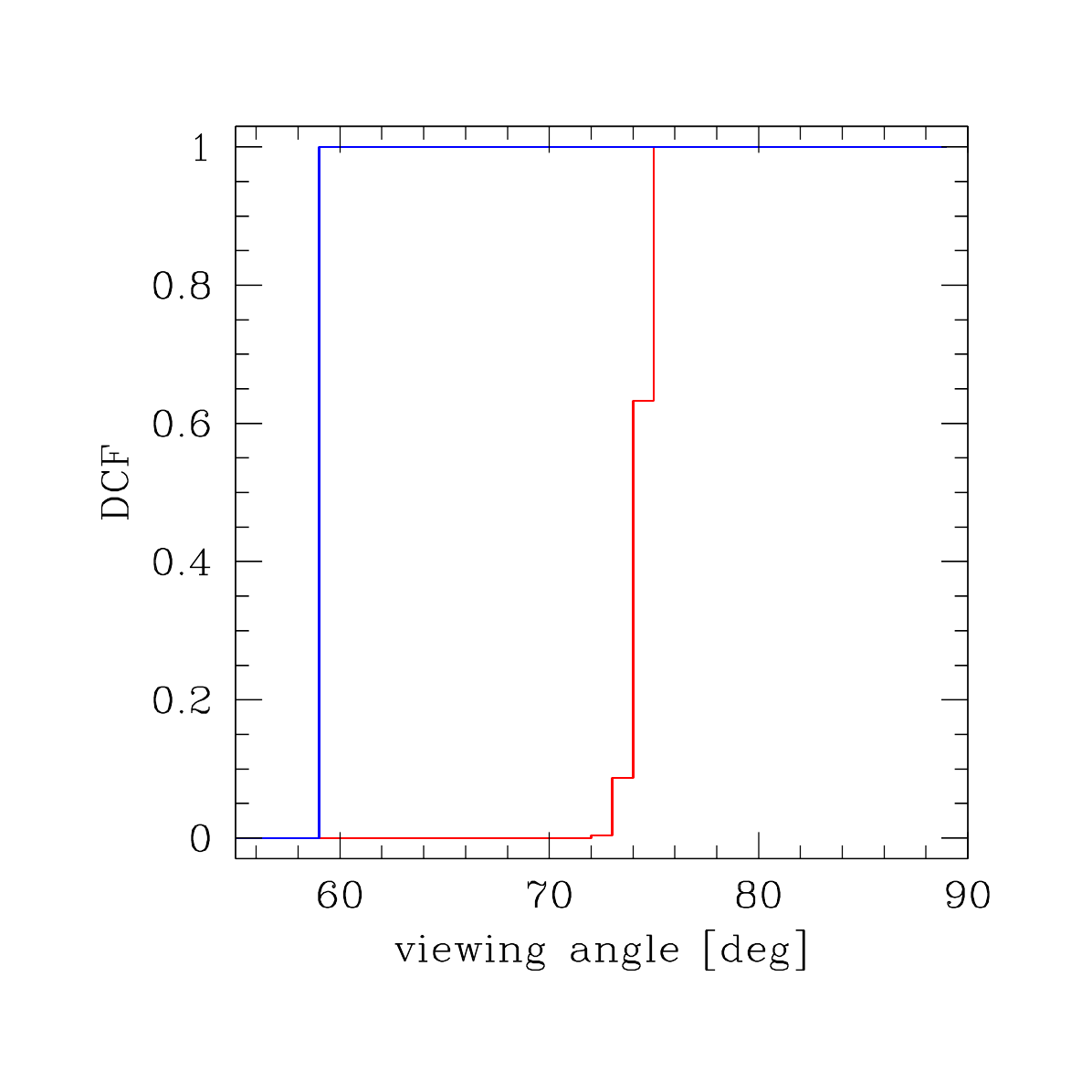}
\caption{Differential covering factor for optically thin (\textbf{left}) and optically thick (\textbf{right}) outflow mechanism. Parameters: black hole mass of $10^{8} M_{\odot}$, dimensionless accretion rate $\dot m = 1$, and two values of the metallicity $Z = 1$ (red line) and $Z = 5$ (blue line) in units of solar metallicity.}\label{fig:differential}
\end{figure}

The DCF(i) for the model in these two outflow efficiencies is shown in Figure~\ref{fig:differential}. In the optically thin launching case, the covering is always small, below 0.02 for the viewing angles below 86 deg. Such a medium is mostly transparent, and the efficiency of production of the BLR lines is far too small, since the expected global covering factor is usually estimated as $\sim$20\%. In the optically thick launching case, the situation is the opposite. The partial coverage is seen only between the viewing angles 72 and 75 deg, and, below this, the BLR intercepts all the radiation. In both cases, there are no clouds along the line of sight below 72 deg, since the dynamics of the cloud motion resulting from the radiation pressure are the same, and its direct impact on reflects the radiation from the extended disk acting on a cloud. Thus, in this picture, for the adopted parameters, clouds at lower inclinations (or, equivalently, lower viewing angles) are not expected. The rise in the value of the DCF with the angle is relatively shallow in the first model. It is not uniform, as, for lower angles, escaping clouds are seen, while, at higher angles, numerous clouds constituting the failed wind appear. In the second case, the saturation of the DCF is rapid since the clouds are so numerous. Thus, in this case, the BLR forms a sharp edge between the cloud-free line of sight and full coverage. If we assume a lower individual cloud density, the transition is even sharper, since the medium becomes less clumpy.  DCF(i) has thus two characteristic points that  conveniently characterize the distribution: $i_{min}$ at which the first cloud appears, and $i_{full}$ when clouds start to cover fully the sky at this and higher viewing angles. We use these quantities when reporting our results in Table~\ref{tab1}.

\begin{figure}[H]
\includegraphics[width=14 cm]{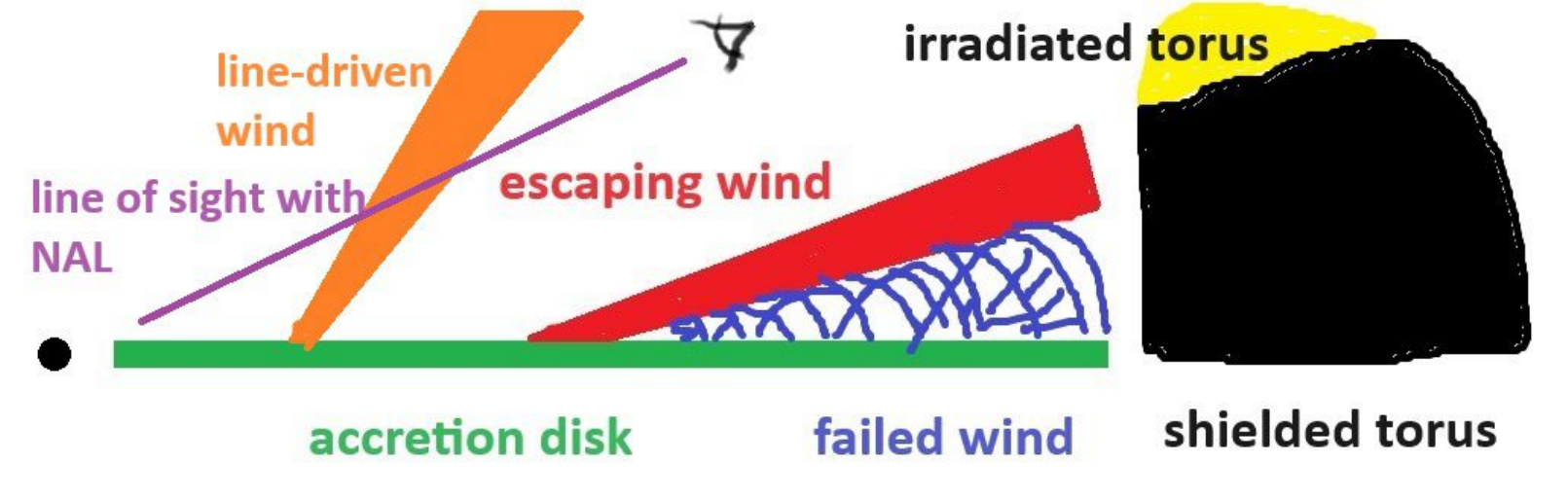}
\caption{Schematic picture of the FRADO flow with an outer torus in thick wind approximation. Irradiation of the outer torus is only in its upper parts, where the central radiation is not intercepted by BLR. An inner line-driven wind and a line of sight showing the narrow absorption lines are also depicted.
\label{fig:schematic}}
\end{figure}

This new computation shows that the dust-driven wind forms complete coverage for the higher viewing angles. In \citet{BAL_2023}, we considered only an escaping stream as a covering medium, and we separately addressed the issue of whether or not the outer dusty/molecular torus was present, as it modified the statistical expectations for the BAL/no-BAL number of sources. Here, we see that the failed part of the wind, which was not included by \citet{BAL_2023}, is actually important in intercepting the radiation from the central regions. In the thick wind scenario, the outer dusty molecular torus is not irradiated at all, unless its angular extension is larger than the wind. We illustrate this schematically in Figure~\ref{fig:schematic}. It may cause a selective dusty outflow from the torus' upper regions, leading to the formation of polar dust as seen in interferometric studies in the IR band (e.g., \citep{Wittkowski2004, garcia2019,Toba2021}). The lower part of the outer torus is not irradiated, and its role is taken by the dusty clouds of the BLR.

The total covering factor for the canonical model in the thick wind option is 0.258, which is quite interesting from the point of view of the expectations based on the line intensities studied by many authors (e.g., \citep{marek2012, elitzur2014}). The case of the optically thin wind gives the covering factor $ 3.9 \times 10^{-3}$, which would give no emission lines irrespective of the spectral shape of the incident continuum. Since the thin/thick winds are extreme approximations, we can consider also an intermediate case, simply assuming the cloud number normalization as a geometrical mean between these two extreme solutions. Such a solution still does not form an optically thick BLR, although the DCF is now $\sim 0.4$, close to the equatorial plane. The global covering factor is still too low, at 0.055, to lead to intense emission lines.

However, as already concluded in \citet{naddaf2021}, the geometry of the flow is strongly dependent on the source parameters (black hole mass, Eddington ratio), as well as on the metallicity, since more heavy metals---and more dust---accelerate the outflow~\citep{Naddaf2022}. We therefore performed similar computations, with the results presented in Table~\ref{tab1}. The increase in metallicity from 1 to 5 times solar did not change qualitatively the model. The whole flow is still optically thin (see Figure ~\ref{fig:differential}), and the global covering factor is still very small, at 0.03, although larger than in case of the solar metalicity, at 0.004. In the thick case, the global covering factor also rises considerably, to 0.51. The DCF rises sharply at 59~deg, with no shallow transition between partial covering and full covering. We also tested the dependence on the local cloud density. The assumption of a fixed column density and lower local density gives a larger size of a single cloud and finally a smaller number of more massive clouds. However, the results for the DFC and $C_{global}$ are practically unaffected. Similarly, the assumption of the lower value of the column density has very little effect on the final results, since, in this case, we require 10 clouds to be present along the line of sight to block the view at a specific viewing angle, but, at the same time, the number of clouds is larger to provide the requested outflow rate.

\begin{table}[H] 
\caption{The parameters of the BLR for $M_{BH}$ of $10^{8} M_{\odot}$ and two values for $\dot m$ and $Z$ each.\label{tab1}}
\newcolumntype{C}{>{\centering\arraybackslash}X}
\begin{tabularx}{\textwidth}{CCCCCC}
\toprule
\boldmath{$\log M_{BH}$}	& \boldmath{$\dot m$}	& \boldmath{$Z$} & \boldmath{$i_{min}$} & \boldmath{$i_{full}$}& \boldmath{$C_{global}$}\\
                        &                   &  \textbf{[}\boldmath{$Z_{\odot}$}\textbf{]}    &  \textbf{deg} & \textbf{deg} &\\
\midrule
8.0	& 1.0		& 1.0 & 72 & 75 & 0.26\\
8.0 & 1.0       & 5.0 &  59 & 59 & 0.51 \\
8.0 & 0.1       & 1.0  & 85 & 85 & 0.087 \\
8.0 & 0.1       & 5.0  & 72 & 73 & 0.28 \\
\bottomrule
\end{tabularx}
\noindent{\footnotesize{
The output parameters ($i_{min}$, $i_{full}$, and $C_{global}$) represent the smallest viewing angle populated with clouds, the viewing angle at which clouds block entirely the line of sight, and the global viewing angle, respectively.}}
\end{table}

The covering factor in our model clearly rises with the Eddington ratio in the thick wind case, since the outflow is more vigorous and covers a larger solid angle. The velocity in the outflow is slightly higher in higher Eddington rate sources but this effect is not strong enough to create a leaky BLR. In the case of an Eddington ratio of 0.1 at solar metallicity, there is no net outflow from the accretion disk, the BLR shows only a failed wind characteristic, and the covering factor is then purely geometrically calculated, as explained in Section~\ref{sect:DCF}.

\section{Discussion}

Our results show that the dust-driven model FRADO of the formation of the BLR is efficient enough to give the expected covering factor. However, this only happens when we assume the optically thick mechanism driving the outflow, in which not only the momentum of the radiation but the energy of the radiation is converted into the outflow. The optically thin case (momentum-driven) leads to the formation of the transparent BLR, which would not produce the emission lines efficiently enough to explain the typical AGN spectra.

Such a slow, optically thick wind is consistent with dusty stellar winds, characteristic of Wolf--Rayet (WR) stars or luminous blue variables (LBVs). Such winds are very difficult to model in great detail, and our approach is only a first step in the direction of creating a full model of the outer part of a luminous AGN. 

Massive outflow is needed to give the required large covering factor, but, at the same time, the outflow may carry a significant amount of material. Only part of the material is in a failed wind, returning to the disk. Part of the material escapes, as already discussed in \citet{naddaf2022pas}, and this flow becomes comparable to the accretion rate when the Eddington rate and/or metallicity is high. In such a case, the disk cannot be treated as a stationary disk with the accretion rate independent of the radius. Such massive outflows (well above a solar mass per year) are seen in many AGNs, but well-resolved winds come from large distances where also stars and stellar winds are present (e.g., \citep{fiore2017,tadhunter2019,davies2023}), and the possible outflow co-spatial with the BLR cannot be well identified. The non-stationarity should be reflected in the spectral shape of the disk (accretion rate dropping inwards should lead to redder spectra than expected from the standard Shakura--Sunyaev model, but this part of the disk emits in the IR, where the contribution from the host and from the circumnuclear dust is important, so the bare disk component is difficult to delineate.

The FRADO model is based on careful computations of the radiation pressure acting on dust, and some predictions are consistent with observationally motivated models, while some are not. The model predicts the stream of escaping material for a range of Eddington rates, high black hole masses, and metallicity, as in the model of \citet{elvis2000}. However, in this model, the line of sight to the nucleus inclined more than the stream of material was relatively unblocked, leading only to the formation of narrow absorption lines (NAL). In our computations, these lines of sight are heavily obscured. It may be that this model mostly applies to the line-driven wind appropriate for the highly ionized part of the BLR, while our model describes the outer, low ionization part. In this case, we would have more zones, as in the scenario from \citet{elvis2010}. Thus, narrow absorption lines could form only at viewing angles smaller than covered by our FRADO model but larger than $\sim$45 deg set by the line-driven wind (see Figure~\ref{fig:schematic}). This leaves a lot of space for such lines in models with a relatively low black hole mass, accretion rate, and metallicity, but less or nothing at the other end of the parameter space. In a recent study of the NAL in a large quasar sample of over 40,000 quasars, the authors identified some absorption features in 43\% of the sources, but finally most of the features were classified as cosmological or environmental absorbers, and only 4.5\% were considered as intrinsic \citep{chen2015}. Thus, perhaps, in bright quasars, such features are indeed rare.

In our model, the covering factor increases with the Eddington luminosity. This is not consistent with the Baldwin effect \citep{baldwin1977}, which shows a decreasing trend of the line equivalent width with the source monochromatic luminosity. However, this effect is mostly seen in highly ionized lines, like CIV, and it is not clear whether the same trend applies to low ionization lines like H$\beta$ or Mg II. In addition, our computation of the covering factor itself does not allow us to conclude on the line emissivity, as this depends also on the incident flux and the local density. Such computations could be performed but they are beyond the scope of the current project.

Since our model is based on the Shakura--Sunyaev disk, it directly applies to sources with an Eddington ratio above a few, where the cold Keplerian accretion disk extends down to the innermost stable circular orbit. The model also requires a stationary disk that extends far out to the distances of a fraction of a parsec. There are no other constraints---it can apply also to radio-loud sources like 3C 273 or flat-spectrum radio quasars (FSRQs). There, the BLR and the dusty torus are present (see, e.g., \citep{prandini2022} for a recent review). However, radio-loud sources likely form a blazar main sequence \citep{cavaliere2002}, with the low luminosity tail formed by BL Lac objects with very weak emission lines. In BL Lacs, the inner radius of the cold disk is distant, or the cold disk is actually absent, and the intrinsic shape of the radiation (not the boosted part coming from the jet) is very different from the one considered in FRADO. The same problem is seen in radio-quiet sources classified as low-ionization nuclear emission-line regions (LINERS). These sources are not well understood, but they clearly do not have a profound cold disk emission (see, e.g., \citep{marquez2017} for a recent discussion). We could consider a modification of the scenario considering a sequence of AGNs with  the Eddington rate decreasing. Initially, when receding the cold disk, only the hot and warm corona would modify the incident radiation seen by the clouds at a large height from the disk, but, at the same time, our launching mechanism becomes less and less efficient, and the low Eddington ratio sources in FRADO simply develop the failed wind resembling the turbulent disk. Direct irradiation becomes less efficient but scattered radiation may have an increasing role. Finally, at very low Eddington rates, the disk region able to launch any dust-driven wind will disappear, replaced by the hot flow, like in Sgr A*. Then, any---even narrow---line emission must come from an illuminated interstellar medium, and FRADO would not apply, even in its modified form. 

A specific question arises as to whether the FRADO model does apply to Changing-Look AGNs.  They can be classified as changing-obscuration objects that show strong variability in the line-of-sight column density, mostly associated with clouds or outflows eclipsing the central engine of the AGN, and changing-state AGNs in which the continuum emission and broad emission lines appear or disappear, and these are typically triggered by strong changes in the accretion rate of the supermassive black hole (see, e.g., \citep{ricci2023}). These sources are typically radiating at a few percent of the Eddington rate (e.g., \citep{CL_stat2022}), so FRADO can apply. However, the fast-changing obscuration can happen only for angles between $i_{min}$ and $i_{full}$, where resolved individual clouds are present and can cause a temporal absorption event. Such short (timescale of a day) obscuration events rarely appear in the X-ray data \citep{risaliti2009,risaliti2011,deMarco2020}, and such events are consistent with expectations from FRADO. However, a single cloud shields only the compact X-ray source and not the entire BLR. Large changes like multiple events of the obscuration of the BLR (see, e.g., \citep{Mao2022} for the source NGC 3227) are likely caused by a stream of material ejected closer to the central black hole than the BLR itself, or at its inner edge but in a non-stationary way.
The estimation of the distance from the ionization state of the material does not give a clear answer; some estimates give rather small distances of the order of 100 $R_g$ (for the source NGC 1365 \citep{risaliti2005}).  Thus, large CL events might be rather intrinsic, with the innermost part of the flow affected by the rapid transition between the cold disk and the hot flow (e.g., \citep{sniegowska2020,Lyu_Wu2022}). Such a rapid transition implies the development of hot and warm corona, not included in our model, and additionally this also means a departure from the stationary flow, as the accretion rate at larger distances cannot change in the timescale of months/years. This could be treated within FRADO, but only after considerable modification of the developed computer program.

\section{Conclusions}

In this paper, we calculated the differential and global covering factors of broad-line region clouds that are produced due to the radiatively dust-driven mechanism described by the FRADO model. The calculations were performed for the canonical value of a black hole mass of $10^{8} M_{\odot}$ at two different accretion rates of 1 and 0.1 Eddington and metallicities of 1 and 5~times solar. We found that the differential covering factor in general does not saturate if the optically thin approximation for the clouds is adopted; however, in the optically thick regime, a sharp spike to one in saturation was found. The results were shown to be directly connected to two among the main three parameters of the source, i.e., the accretion rate and metallicity of the clouds. We found that as these global parameters increased, the global covering factor also increased. Moreover, according to previous studies on the shape and geometry of the BLR \citep{naddaf2021, Naddaf2022}, we also expect the covering factor to highly likely follow the change in the black hole mass as well, correspondingly. We will address the dependence of the covering factor on the source parameters in detail in the future for a significantly larger and denser set of initial parameters.

\vspace{6pt}
\authorcontributions{All authors have read and agreed to the published version of the manuscript.} 

\funding{The project was partially supported by the Polish Funding Agency National Science Centre, project 2017/26/A/ST9/00756 (MAESTRO 9). B.C. acknowledges the Czech--Polish mobility program (M\v{S}MT 8J20PL037 and PPN/BCZ/2019/1/00069). This project received funding from the European Research Council (ERC) under the European Union’s Horizon 2020 research and innovation program (grant agreement No. [951549]). In addition, this research was supported by the University of Liege under Special Funds for Research, IPD-STEMA Program.}

\dataavailability{Data are contained within the article.}


\conflictsofinterest{The authors declare no conflicts of interest.}

\begin{adjustwidth}{-\extralength}{0cm}
\reftitle{References}


\externalbibliography{yes}




\PublishersNote{}
\end{adjustwidth}
\end{document}